\documentclass[letterpaper,10pt]{article}

\usepackage[utf8]{inputenc}
\usepackage{amsmath}
\usepackage{amssymb}

\newcommand{\dd}{\textrm{d}}
\newcommand{\im}{{\mathbb{I}}{\mathrm{m}}}

\newcommand{\hypergf}{{}_{2}F_{1}}

\title{Electromagnetic quasinormal modes of five-dimensional topological black holes}

\author{D.\ V.\ G\'omez-Navarro, A.\ L\'opez-Ortega\thanks{alopezo@ipn.mx} \\ 
Departamento de F\'{\i}sica. \\
Escuela Superior de F\'{\i}sica y Matem\'aticas. \\
Instituto Polit\'ecnico Nacional. \\
Unidad Profesional Adolfo L\'opez Mateos. Edificio 9. \\
M\'exico, D.\ F., M\'exico. \\
C.\ P.\ 07738 }

\begin{document}

\maketitle

\begin{abstract}

We calculate exactly the QNF of the vector  type and scalar type electromagnetic fields propagating on a family of five-dimensional topological black holes. To get a discrete spectrum of quasinormal frequencies for the scalar type electromagnetic field we find that it is necessary to change the boundary condition usually imposed at the asymptotic region. Furthermore for the vector type electromagnetic field we impose the usual boundary condition at the asymptotic region and we discuss the existence of unstable quasinormal modes in the five-dimensional topological black holes.



\end{abstract}

\section{Introduction}
\label{s: Introduction}

The quasinormal modes (QNM) are the characteristic oscillations of a test field or of the metric perturbations that satisfy the appropriate boundary conditions near the event horizon and at the asymptotic region of the black hole \cite{Kokkotas:1999bd}, \cite{Berti:2009kk}. Thus the QNM appear when the black hole is perturbed from its equilibrium state and they depend on the physical properties of the black hole and the field. Recently the QNM have found many applications in different studies about the classical stability of the black holes \cite{Kokkotas:1999bd}, \cite{Berti:2009kk}, the area spectrum of the event horizon \cite{Hod:1998vk}, \cite{Maggiore:2007nq}, and motivated by the AdS-CFT correspondence  \cite{Aharony:1999ti}, the QNM spectra of asymptotically anti-de Sitter black holes are extensively studied since they are useful in the calculation of the decay rates in dual theories \cite{Aharony:1999ti}--\cite{LopezOrtega:2010uu}. 

Furthermore the exactly solvable systems are relevant in physics, since in these systems we can explore in detail their physical properties. Thus we think that in the research line of black hole perturbations is useful the search and study of exactly solvable systems.  In particular for asymptotically anti-de Sitter black holes it is possible to calculate exactly the quasinormal frequencies (QNF) of several fields \cite{Cardoso:2001hn}--\cite{Becar:2012bj}. We know that this is the case for the BTZ black holes \cite{Cardoso:2001hn}--\cite{Birmingham:2001pj}, the massless topological black holes \cite{Aros:2002te}--\cite{LopezOrtega:2010uu}, and the five-dimensional topological black holes \cite{Becar:2012bj} of Refs.\ \cite{Cai:1998vy}--\cite{Cai:2009ar}. See also \cite{Gonzalez:2010vv}--\cite{Oliva:2010xn} for related examples of exact determination of  QNF.  

The QNF spectrum of the electromagnetic field in asymptotically anti-de Sitter black holes has been explored previously \cite{Cardoso:2001bb}--\cite{LopezOrtega:2006vn},  \cite{Nunez:2003eq}--\cite{Miranda:2008vb}. Among the motivations we find that the electromagnetic field behaves in a different way than other classical fields and its analysis is physically more relevant than the study of the Klein-Gordon field.  Additionally in the AdS-CFT correspondence the QNF of the electromagnetic field in asymptotically anti-de Sitter spacetimes are related to the poles of the retarded Green functions of the R-symmetry currents.  Thus for the electromagnetic field we believe that it is convenient to determine its spectrum of QNF in asymptotically anti-de Sitter black holes.

Based on the results by Kodama and Ishibashi on the simplification of the Maxwell equations in static spacetimes \cite{Kodama:2003kk}, (see also \cite{Lopez-Ortega:2014oha}) here we generalize previous results on the QNM of the Klein-Gordon field \cite{Becar:2012bj} in the five-dimensional topological black holes of Refs.\ \cite{Cai:1998vy}--\cite{Cai:2009ar} and calculate exactly the spectrum of QNF for the electromagnetic field. Using these results we study the stability of the topological black holes of Refs.\ \cite{Cai:1998vy}--\cite{Cai:2009ar} under perturbations.  At this point it is convenient to mention that we discuss the possible existence of unstable QNM for the electromagnetic field for some values of the parameters. 

We organize this work as follows. Following Kodama and Ishibashi \cite{Kodama:2003kk} (see also \cite{Lopez-Ortega:2014oha}) in Sect.\ \ref{s: Maxwell equations} we recall some relevant results about the simplification of the vacuum Maxwell equations in static spacetimes to two differential equations, one for the vector type electromagnetic field and another for the scalar type electromagnetic field. In Sect.\ \ref{s: black holes} we enumerate the relevant features of the five-dimensional topological black holes of Refs.\ \cite{Cai:1998vy}--\cite{Cai:2009ar} that we study in this work. In Sect.\ \ref{s: vector type} we calculate exactly the QNF of the vector type electromagnetic field propagating on these five-dimensional topological black holes and discuss the stability of the QNM. We make a similar calculation for the scalar type electromagnetic field in Sect.\ \ref{s: scalar type}, but in this case we need to make a careful study of the boundary condition at the asymptotic region since the usual boundary condition leads to a continuum of QNF and we modify the boundary condition to get a discrete set of QNF that depends on the parameters of the black hole and the field. Finally we discuss the main results in Sect.\ \ref{s: Discussion}.

\section{Maxwell equations}
\label{s: Maxwell equations}

As is well known, we can write the line element of a $D$-dimensional generalization of the spherically symmetric spacetime in the form \cite{Kodama:2003kk}
\begin{equation} \label{e: metric maximally symmetric}
 \dd s^2 = g_{a b} (x) \dd x^a \dd x^b + r^2 (x) \dd \Omega_{D-2}^2,
\end{equation} 
where $a,b = 1,2,$ $ \dd \Omega_{D-2}^2 = \hat{\gamma}_{ij} \dd \hat{y}^i \dd \hat{y}^j $, $i,j=1,2,\dots,D-2$, is the line element of the $(D-2)$-dimensional maximally symmetric base manifold with metric $\hat{\gamma}_{ij}$ and whose Ricci tensor fulfills $\hat{R}_{ij} = (D-3)K \hat{\gamma}_{ij}$, that is, the base manifolds are of Einstein type. Here $K$ is a constant determined by the scalar curvature of the base manifold and can be normalized to the values $K=0,\pm 1$ \cite{Kodama:2003kk}. In what follows we assume that the bidimensional line element that appears in the metric (\ref{e: metric maximally symmetric}) is given by
\begin{equation} \label{e: bidimensional metric}
 \dd s^2_{2} = g_{ab} (x) \dd x^a \dd x^b= -\mathcal{F} \dd t^2 + \frac{\dd r^2}{\mathcal{G}},
\end{equation} 
with $\mathcal{F}$ and $\mathcal{G}$ functions of the radial coordinate $r$.

If we denote the Maxwell tensor by $F_{\mu \nu}$ then the Maxwell equations in vacuum are 
\begin{equation}
 \nabla_{[\sigma} F_{\mu \nu]} = 0, \qquad \qquad \qquad \nabla_\mu F^{\mu \nu} = 0.
\end{equation}
It is well known that if we make a harmonic sum on the scalar and vector eigenfunctions of the Laplacian on the base manifold  $\dd \Omega_{D-2}^2$, the Maxwell equations  in a spacetime of the form (\ref{e: metric maximally symmetric}) simplify to \cite{Kodama:2003kk}, \cite{Lopez-Ortega:2014oha} 
\begin{align} \label{e: master vector type}
 D_a D^a  \Phi_V &- \frac{D-4}{4r} \frac{\dd \mathcal{G}}{\dd r}  \Phi_V - \frac{(D-4)(D-6) \mathcal{G}}{4 r^2}  \Phi_V \nonumber \\ 
 & - \frac{D-4}{4r} \frac{\mathcal{G}}{\mathcal{F}} \frac{\dd \mathcal{F}}{\dd r} \Phi_V - \frac{k_V^2 + (D-3)K}{r^2} \Phi_V= 0,  
\end{align}
for the vector type electromagnetic field and 
\begin{align} \label{e: master scalar type}
 D_a D^a \Phi_S &- \frac{(D-2)(D-4)}{4} \frac{\mathcal{G}}{r^2} \Phi_S + \frac{\dd \mathcal{G}}{\dd r} \frac{D-4}{4r} \Phi_S \nonumber \\ 
 &+ \frac{\mathcal{G}}{r \mathcal{F}} \frac{\dd \mathcal{F}}{\dd r} \frac{D-4}{4} \Phi_S -\frac{k_S^2}{r^2} \Phi_S = 0 ,
\end{align}
for the scalar type electromagnetic field. Here the symbol $D_a$ denotes the covariant derivative for the bidimensional metric $g_{ab}(x)$, the functions $\Phi_V$ and $\Phi_S$ depend on the coordinates  $x^a$  of the two-dimensional space with metric $g_{ab}$ and they contain the relevant information about the dynamics of the vector type and scalar type electromagnetic fields in spacetimes of the form (\ref{e: metric maximally symmetric}).  In the previous formulas $k_V^2$ ($k_S^2$) are the eigenvalues of the vector harmonics $\mathbb{V}_i$ (scalar harmonics $\mathbb{S}$)  on the maximally symmetric base manifold with line element $\dd \Omega_{D-2}^2$, that is, they satisfy \cite{Kodama:2003kk}
\begin{equation}
(\hat{D}_i \hat{D}^i + k_V^2)  \mathbb{V}_j = 0, \qquad \qquad \hat{D}^i  \mathbb{V}_i = 0, \qquad \qquad ((\hat{D}_i \hat{D}^i + k_S^2)  \mathbb{S} = 0 ),
\end{equation} 
where $\hat{D}_i$ is the covariant derivative on the maximally symmetric base manifold. For $\mathcal{F}= \mathcal{G} = f$ we point out that in Eqs.\ (\ref{e: master vector type}) and (\ref{e: master scalar type}) the operator $D_a D^a$ takes the form
\begin{equation} \label{e: laplacian bidimensional}
 D_a D^a = -\frac{1}{f} \partial_t^2 + \partial_r (f \partial_r).
\end{equation}

\section{Five dimensional topological black holes}
\label{s: black holes}

The five-dimensional topological black holes that we study in this work have the line element
\begin{equation} \label{e: 5D black holes}
  \dd s^2 = -f \dd t^2 + \frac{\dd r^2}{f} + r^2 \dd \Omega_{K}^2 ,
\end{equation} 
where $\dd \Omega_{K}^2$ is the line element of the three-dimensional maximally symmetric base manifold and the function $f$ takes the form \cite{Cai:1998vy,Cai:2001dz,Cai:2009ar} 
\begin{equation}
 f = - \frac{\Lambda}{3} r^2 + K \pm \sqrt{c_0},
\end{equation} 
where $\Lambda$ is a negative constant, $K=0, \pm 1$, and $c_0$ is a non negative constant. We notice that for the three-dimensional base manifold the scalar curvature is equal to $6 K$. The solution with positive sign of $\sqrt{c_0}$ is usually called the plus branch, whereas the solution with negative sign of $\sqrt{c_0}$  is usually known as the minus branch  \cite{Cai:2009ar}. The topological black holes (\ref{e: 5D black holes}) are solutions of several gravity theories as the five-dimensional Chern-Simmons theory \cite{Cai:1998vy}, the five-dimensional Gauss-Bonnet gravity with special Gauss-Bonnet coefficient \cite{Cai:2001dz}, and the five-dimensional $z=4$ Ho\v rava-Lifshitz gravity \cite{Cai:2009ar}.  

The event horizon of the topological black holes (\ref{e: 5D black holes}) is determined by 
\begin{equation}
 r_+^2 = -\frac{3}{\Lambda} (\mp \sqrt{c_0} - K).
\end{equation} 
Since we impose that the radius of the event horizon is positive, we need that  the parameters $c_0$ and $K$ fulfill 
\begin{equation}
 \mp \sqrt{c_0} - K > 0.
\end{equation} 
Notice that depending on the value of $K$ we have one or two positive values of the event horizon according to the following list.  
\begin{enumerate}
 \item For $K=0$ and $K=1$ the black hole exists in the minus branch.
 \item For $K=-1$ to have a black hole in the plus branch we require that $\sqrt{c_0} < 1$.
 \item For $K=-1$ in the minus branch we always have a black hole.
\end{enumerate}
Thus for $K=-1$ we have black holes in the two branches of the solution (\ref{e: 5D black holes}) \cite{Cai:2009ar}. 

In the following sections it is useful to define the quantity 
\begin{equation} \label{e: p definition}
 p = K \pm \sqrt{c_0},
\end{equation} 
and if the black hole exists, then $p < 0$. We also notice that for $K=-1$ the parameter $p$ satisfies $|p| > 1$ in the minus branch and $|p|<1$ in the plus branch.

\section{Quasinormal frequencies of the vector type electromagnetic field}
\label{s: vector type}

To extend the previous results of \cite{Becar:2012bj} on the QNM spectrum of the five-di\-men\-sion\-al topological black holes (\ref{e: 5D black holes}) in what follows we calculate exactly the QNF of the electromagnetic field moving on these black holes. Following Refs.\ \cite{Cardoso:2001hn}--\cite{Becar:2012bj} for the topological black holes (\ref{e: 5D black holes}) we define their QNM as the oscillations that satisfy the boundary conditions 
\begin{description}
 \item[i)] The electromagnetic field is purely ingoing near the black hole horizon.
 \item[ii)] The electromagnetic field goes to zero as $r \to + \infty$.
\end{description}

We notice that the line elements of the five-dimensional topological black holes (\ref{e: 5D black holes}) are of the form (\ref{e: metric maximally symmetric}) with $\mathcal{F}= \mathcal{G} = f$, therefore we can use Eqs.\  (\ref{e: master vector type}) and (\ref{e: master scalar type}) to study the propagation of the electromagnetic fields on these backgrounds. Here we begin with the vector type electromagnetic field.

Since the topological black holes (\ref{e: 5D black holes}) are static and taking into account the expression (\ref{e: laplacian bidimensional}) for the operator $D_a D^a$, we propose that the function $\Phi_V$ takes the form 
\begin{equation} \label{e: ansatz vector type}
 \Phi_V = \textrm{e}^{- i \omega t} R_V (r),
\end{equation} 
and therefore from Eq.\  (\ref{e: master vector type}) we obtain that the radial function $R_V$ must be a solution of the differential equation
\begin{equation} \label{e: radial vector type}
 f \frac{\dd^2 R_V}{\dd r^2} + \frac{\dd f}{\dd r} \frac{\dd R_V}{\dd r} + \left(\frac{\omega^2}{f} - \frac{1}{2 r} \frac{\dd f}{\dd r} + \frac{f}{4 r^ 2}  - \frac{k_V^2 + 2 K}{r^2} \right) R_V =0 .
\end{equation} 
To  solve exactly the previous differential equation, as in Ref.\  \cite{Becar:2012bj},  it is convenient to define the variable\footnote{Notice that the quantity $v$ varies over the range $0<v<1$.} 
\begin{equation} \label{e: v definition}
 v = 1 - \frac{3 p}{\Lambda r^2},
\end{equation} 
with $p$ already given in the formula (\ref{e: p definition}). Using the variable $v$ we find that Eq.\  (\ref{e: radial vector type}) transforms into 
\begin{equation}
 \frac{\dd^2 R_V}{\dd v^2} +\left( \frac{1}{v} - \frac{1/2}{1-v} \right) \frac{\dd R_V}{\dd v} + \left(\frac{F +G - 3/16}{v(1-v)} + \frac{F}{v^2} - \frac{3/16}{(1-v)^2} \right) R_V =0,
\end{equation} 
where we define the constants $F$ and $G$ by
\begin{equation} \label{e: F G definition}
 F = \frac{3 \omega^2}{4 p \Lambda}, \qquad \qquad \qquad G = \frac{k_V^2 + 2K}{4 p} - \frac{1}{16}.
\end{equation} 

Taking the function $R_V$ in the form
\begin{equation}
 R_V = (1-v)^{A_V} v^{B_V} R_{2V},
\end{equation} 
with the parameters $A_V$ and $B_V$ being solutions to the algebraic equations 
\begin{equation}
 A_V^2 - \frac{A_V}{2} - \frac{3}{16} = 0, \qquad \qquad \qquad B_V^2 + F = 0,
\end{equation} 
we find that the function $R_{2V}$ is a solution of the differential equation 
\begin{align}
v (1-v) \frac{\dd^2 R_{2V}}{\dd v^2} &+(2 B_V + 1 -(2 B_V + 2 A_V + 3/2)v)\frac{\dd R_{2V} }{\dd v} \nonumber \\
&- (2 A_V B_V + B_V/2 + A_V + 3/16 - F -G)  R_{2V} = 0 .
\end{align} 
This equation is of hypergeometric type \cite{Abramowitz-book}--\cite{NIST-book} 
\begin{equation} \label{e: hypergeometric equation}
 v(1-v)\frac{\dd^2 h}{\dd v^2} + (\gamma -(\alpha +\beta +1)v)\frac{\dd h}{\dd v} - \alpha \beta h = 0 ,
\end{equation} 
with parameters $\alpha_V$, $\beta_V$, $\gamma_V$ equal to
\begin{align} \label{e: alpha beta gamma vector type}
 \alpha_V &= A_V + B_V + \tfrac{1}{4} + \tfrac{1}{2}\sqrt{\tfrac{1}{4}+4G}, \quad  \beta_V = A_V + B_V + \tfrac{1}{4} - \tfrac{1}{2}\sqrt{\tfrac{1}{4}+4G}, \nonumber \\ 
 \gamma_V &= 2 B_V + 1.
\end{align} 

In what follows we choose
\begin{equation} \label{e: A B vector type}
 A_V = \frac{3}{4}, \qquad \qquad \qquad B_V = \frac{i \omega q}{2},
\end{equation} 
where
\begin{equation}
 q = \sqrt{\frac{3}{p \Lambda}}.
\end{equation} 
Notice that $q > 0$. For these values of $A_V$, $B_V$ we get that $\gamma_V-\alpha_V-\beta_V=-1$.

From these results we find that the function $R_V$ is equal to \cite{Abramowitz-book}--\cite{NIST-book}
\begin{align} \label{e: solution vector type}
 R_V=(1-v)^{3/4} v^{i \omega q/2} & \left(K_1 \,\,\,\, \hypergf (\alpha_V, \beta_V;\gamma_V;v) \right. \\
 &+ \left. K_2 v^{1-\gamma_V} \hypergf (\alpha_V - \gamma_V +1,\beta_V - \gamma_V +1; 2 -\gamma_V; v ) \right) , \nonumber 
\end{align}
where $K_1$, $K_2$ are constants and $\hypergf (\alpha, \beta;\gamma;v)$ denotes the hypergeometric function \cite{Abramowitz-book}--\cite{NIST-book}. Taking into account that $v \approx 0$ near the horizon of the black hole, from the expression (\ref{e: solution vector type}) for $R_V$, we observe that near the horizon this function behaves as
\begin{equation} \label{e: near horizon vector type}
 R_V \approx K_1 v^{i \omega q / 2} + K_2 v^{- i \omega q / 2} \approx K_1 \textrm{e}^{i \omega r_*} +  K_2 \textrm{e}^{-i \omega r_*} ,
\end{equation} 
where $r_*$ denotes the tortoise coordinate of the five-dimensional topological black hole (\ref{e: 5D black holes})
\begin{equation}
 r_* = \frac{q}{2} \ln \left|\frac{\sqrt{1-v}-1}{\sqrt{1-v}+1} \right|. 
\end{equation} 
Notice that $r_* \to -\infty$ near the horizon and $r_* \to 0$ as $r \to + \infty$.

Since we take a time dependence of the form $\exp(-i \omega t)$ (see the expression (\ref{e: ansatz vector type})) we find that the first term of the expression (\ref{e: near horizon vector type}) is an outgoing wave near the horizon, whereas the second term is an ingoing wave near the horizon. Thus to satisfy the boundary condition i) of the QNM we must take $K_1=0$ and therefore the function $R_V$ fulfilling the boundary condition near the horizon takes the form
\begin{align} \label{e: solution vector type i)}
 R_V &= K_2 (1-v)^{3/4} v^{-i \omega q/2} \hypergf (\alpha_V - \gamma_V +1,\beta_V - \gamma_V +1; 2 -\gamma_V; v ) \nonumber \\
 &=  K_2 (1-v)^{3/4} v^{-i \omega q/2} \hypergf (\hat{\alpha}_V, \hat{\beta}_V;\hat{\gamma}_V;v),
\end{align} 
with 
\begin{equation}
 \hat{\alpha}_V = \alpha_V - \gamma_V +1, \qquad \hat{\beta}_V = \beta_V - \gamma_V +1, \qquad \hat{\gamma}_V = 2 -\gamma_V.
\end{equation} 

To study the behavior of the field as $r \to + \infty$ ($v \to 1$) the usual procedure for many exactly solvable problems  \cite{Cardoso:2001hn}--\cite{Becar:2012bj} is to use the Kummer property of the hypergeometric function \cite{Abramowitz-book}--\cite{NIST-book}, but for the vector type field, since the parameters of the hypergeometric function that appears in the solution (\ref{e: solution vector type i)}) fulfill $\hat{\gamma}_V - \hat{\beta}_V - \hat{\alpha}_V = -1 $ and the Kummer property is not valid when the values of the parameters satisfy this condition \cite{Abramowitz-book}--\cite{NIST-book}, we cannot employ the usual procedure. Nevertheless for $\gamma - \alpha - \beta = -m $, $m=0,1,2,\dots$, the hypergeometric function satisfies 
\begin{align} \label{e: hypergeometric property integer}
{}_2F_1 (\alpha,\beta;\gamma;v) &=   \frac{\Gamma(\gamma) \Gamma(m)}{\Gamma(\alpha)\Gamma(\beta)} (1-v)^{-m} \sum_{s=0}^{m-1}\frac{(\alpha-m)_s (\beta-m)_s}{s! (1-m)_s}(1-v)^s \\ 
&  + \frac{(-1)^{m+1} \Gamma(\gamma)}{\Gamma(\alpha-m)\Gamma(\beta-m)} \sum_{s=0}^\infty \frac{(\alpha)_s (\beta)_s}{s!(m+s)!}(1-v)^s  \nonumber \\
&\times [\ln(1-v) -\psi(s+1) -\psi(s+m+1)  +\psi(\alpha+s)+\psi(\beta+s)],  \nonumber
\end{align}
when the parameters $\alpha$ and $\beta$ are different from negative integers \cite{Abramowitz-book}--\cite{NIST-book}. In the previous formula $\psi$ is the logarithmic derivative of the gamma function and $(m)_s$ is the Pochhammer symbol. For $m=0$ we must delete the finite sum. Notice that  for the vector type electromagnetic field we have $m=1$.

To analyze the behavior of the field as $r \to + \infty$ we use the property (\ref{e: hypergeometric property integer}) to write the radial function (\ref{e: solution vector type i)}) as 
\begin{align}
 R_V &= K_2 (1-v)^{3/4} v^{-i \omega q/2} \left( \frac{\Gamma(\hat{\gamma}_V) }{\Gamma(\hat{\alpha}_V)\Gamma(\hat{\beta}_V)} \frac{1}{1-v}    \right. \nonumber \\
 &+ \frac{ \Gamma(\hat{\gamma}_V)}{\Gamma(\hat{\alpha}_V-1)\Gamma(\hat{\beta}_V -1)} \sum_{s=0}^\infty \frac{(\hat{\alpha}_V)_s (\hat{\beta}_V)_s}{s!(s+1)!} (1-v)^s  \nonumber \\
 &\times  \left.  
 [\ln(1-v) -\psi(s+1) -\psi(s+2)  +\psi(\hat{\alpha}_V+s)+\psi(\hat{\beta}_V+s)] \right) .
\end{align}
From this expression we obtain that the second factor goes to zero as $r \to + \infty$, but in this limit the first factor behaves in the form
\begin{equation}
 \frac{\Gamma(\hat{\gamma}_V) }{\Gamma(\hat{\alpha}_V)\Gamma(\hat{\beta}_V)} \frac{1}{(1-v)^{1/4}} ,
\end{equation} 
and therefore diverges as $v \to 1$. In a similar way to previous works \cite{Cardoso:2001hn}--\cite{Becar:2012bj}, \cite{Lopez-Ortega:2014oha}, to cancel this term we like to impose the conditions 
\begin{equation} \label{e: conditions vector type}
 \hat{\alpha}_V = -n, \qquad \qquad \qquad \hat{\beta}_V = -n,
\end{equation} 
with $n=0,1,2,\dots$ But if we impose the conditions (\ref{e: conditions vector type}), then we contradict the assumptions under which the property (\ref{e: hypergeometric property integer}) is true. Thus if we use the property (\ref{e: hypergeometric property integer}) then we cannot impose the conditions (\ref{e: conditions vector type}) because we contradict the assumptions under which this property is valid.

Another path is to impose the conditions (\ref{e: conditions vector type}) on the parameters of the solution (\ref{e: solution vector type i)}) and see whether the radial function $R_V$  satisfies the boundary condition ii) of the QNM \cite{Lopez-Ortega:2014oha}. Thus assuming that $\hat{\alpha}_V = -n$, we find that 
\begin{align} \label{e: solution vector polynomial}
 R_V &= K_2 (1-v)^{3/4} v^{-i \omega q/2} \hypergf (-n,\hat{\beta}_V;\hat{\gamma}_V;v )  \\
 &= K_2 (1-v)^{3/4} v^{-i \omega q/2} \frac{(1-\beta_V)_n}{(2-\gamma_V)_n}  \hypergf (-n,\beta_V-\gamma_V+1;\beta_V-n;1-v), \nonumber
\end{align} 
where we use that the hypergeometric function fulfills \cite{NIST-book}
\begin{equation} \label{e: hypergeometric polynomial}
 {}_2F_1(-n,\beta;\gamma;v) = \frac{(\gamma - \beta)_n}{(\gamma)_n} {}_2F_1(-n,\beta;\beta-\gamma-n+1;1-v) .
\end{equation} 
In a straightforward way we obtain that near $v=1$ the function $R_V$ of the expression (\ref{e: solution vector polynomial}) behaves as
\begin{equation}
 R_V \approx (1-v)^{3/4},
\end{equation} 
and therefore it satisfies the boundary condition ii) of the QNM as $v \to 1$. A similar thing happens for $\hat{\beta}_V = -n$. Thus imposing the conditions (\ref{e: conditions vector type}) we obtain that the radial function $R_V$ satisfies the boundary condition ii) of the QNM.

From the conditions (\ref{e: conditions vector type}) and taking into account the values of the parameters (\ref{e: alpha beta gamma vector type}) and (\ref{e: A B vector type}) we get that the QNF of the vector type electromagnetic field are equal to
\begin{equation} \label{e: QNF vector type}
 \omega_V = \pm \frac{1}{q} \sqrt{\frac{k_V^2 + 2 K}{|p|}} -i \frac{2}{q} (n + 1) .
\end{equation} 
It is convenient to notice that the QNF (\ref{e: QNF vector type}) of the vector type electromagnetic field  depend on the value of the parameter $K$ related to the scalar curvature of the base manifold. We  recall that $p < 0$ to have a positive radius of the event horizon. Furthermore we remember that the eigenvalues $k_V^2$ are non negative and they are discrete for $K=1$, whereas they are continuous for $K \leq 0$ \cite{Kodama:2003kk}. From the values (\ref{e: QNF vector type}) for the QNF we get that for $K=0$ and $K=1$ the real part of the QNF for the vector type field is different from zero and the imaginary part is negative, that is, the QNM decay in time since we have a time dependence $\exp(-i \omega t)$, and therefore they are stable for $K=0$ and $K=1$. We remark that for $K=0$ and $K=1$ the QNF (\ref{e: QNF vector type}) are complex.

For $K=-1$ we obtain that the QNF (\ref{e: QNF vector type}) simplify to
\begin{equation} \label{e: QNF vector type K negative}
 \omega_V = \pm \frac{1}{q} \sqrt{\frac{k_V^2 - 2 }{|p|}} - i \frac{2}{q} (n + 1) .
\end{equation} 
Thus for $k_V^2  > 2$ the QNF (\ref{e: QNF vector type K negative}) have real and imaginary parts different from zero and they are stable since $\im (\omega_V) < 0$. Nevertheless if the base manifold allows eigenvalues of the vector harmonics satisfying $0 < k_V^2 < 2$, we obtain that the QNF (\ref{e: QNF vector type K negative}) transform into 
\begin{equation} \label{e: QNF vector type K negative smaller two}
 \omega_V = \pm \frac{i}{q} \sqrt{ \left| \frac{k_V^2 - 2 }{p} \right|} - i \frac{2}{q} (n + 1) ,
\end{equation} 
that is, they are purely imaginary. From the expressions  (\ref{e: QNF vector type K negative}) and (\ref{e: QNF vector type K negative smaller two}) we remark that for $K=-1$ we get complex or purely imaginary QNF depending on the value of the eigenvalues $k_V^2$.  The QNF (\ref{e: QNF vector type K negative smaller two}) with the minus sign of the square root are stable since $\im (\omega_V) < 0$, but the QNF with the plus sign of the square root are unstable if
\begin{equation} \label{e: inequality unstable vector}
 \left| \frac{k_V^2 - 2 }{p} \right| > 4 (n + 1)^2 .
\end{equation} 
We notice that this inequality is not satisfied for $|p| > 1$, but for $k_V^2$ sufficiently small, we can fulfill the inequality (\ref{e: inequality unstable vector}) for  sufficiently small $|p| < 1$, that is, for the plus branch of the topological black holes (\ref{e: 5D black holes}) with $K=-1$. Thus, for these topological black holes the fundamental mode ($n=0$) is unstable for
\begin{equation}
 2 - 4 |p| >  k_V^2 .
\end{equation} 
For the overtones we find that the condition (\ref{e: inequality unstable vector}) becomes
\begin{equation}
 2 - 4 |p| (n+1)^2 > k_V^2 .
\end{equation} 
Thus for the topological black holes (\ref{e: 5D black holes}) with $K=-1$ in the plus branch with sufficiently small values of the parameter $|p|$ we find unstable QNM for the vector type electromagnetic field if the base manifold has eigenvalues of the vector harmonics in the range $0 < k_V^2 < 2$.

Nevertheless, for three-dimensional closed base manifolds and three-di\-men\-sion\-al open base manifolds such that the quantity $ \mathbb{V}^j \hat{D}_i  \mathbb{V}_j$ fall off sufficiently rapidly at infinity, it is known that the eigenvalues of the vector harmonics satisfy $k_V^2 \geq 2 |K|$ \cite{Kodama:2003kk}. Thus for these three-dimensional base manifolds there is no instability of the QNM of the vector type electromagnetic field. We do not know an example of a three-dimensional base Einstein manifold with eigenvalues in the range $0 < k_V^2 < 2$.

\section{Quasinormal frequencies of the scalar type electromagnetic field}
\label{s: scalar type}

Here we extend the results of the previous section and calculate exactly the QNF of the scalar type electromagnetic field propagating on the topological black holes (\ref{e: 5D black holes}). In a similar way to the previous section  we take
\begin{equation}
 \Phi_S = \textrm{e}^{-i \omega t} R_S (r),
\end{equation} 
and from Eq.\ (\ref{e: master scalar type}) we obtain that the function $R_S$ must be a solution of the differential equation
\begin{equation} \label{e: radial scalar type}
 f \frac{\dd^2 R_S}{\dd r^2} + \frac{\dd f}{\dd r} \frac{\dd R_S}{\dd r} + \left(\frac{\omega^2}{f} + \frac{1}{ 2r} \frac{\dd f}{\dd r} - \frac{3f}{4 r^ 2}  - \frac{k_S^2 }{r^2} \right) R_S =0 .
\end{equation} 
To solve exactly this equation we use the variable $v$ defined in the formula (\ref{e: v definition}) to find that it transforms into
\begin{equation}  \label{e: radial scalar type v variable}
 \frac{\dd^2 R_S}{\dd v^2} +\left( \frac{1}{v} - \frac{1/2}{1-v} \right) \frac{\dd R_S}{\dd v} + \left(\frac{F + H + 1/16}{v(1-v)} + \frac{F}{v^2} + \frac{1/16}{(1-v)^2} \right) R_S =0,
\end{equation} 
with $F$ defined in the expression (\ref{e: F G definition}) and 
\begin{equation}
 H = \frac{k_S^2}{4 p} + \frac{3}{16}.
\end{equation} 

To solve Eq.\ (\ref{e: radial scalar type v variable}) we propose that the radial function $R_S$ takes the form 
\begin{equation}
 R_S = (1 -v )^{A_S} v^{B_S} R_{2 S},
\end{equation} 
with the quantities $A_S$ and $B_S$ being solutions of the algebraic equations
\begin{equation}
 A_S^2 - \frac{A_S}{2} + \frac{1}{16} = 0, \qquad \qquad \qquad B_S^2 + F = 0 ,
\end{equation} 
to find that the function $R_{2S}$ is a solution of the differential equation
\begin{align}
v (1-v) \frac{\dd^2 R_{2S}}{\dd v^2} &+(2 B_S+ 1 -(2 B_S + 2 A_S + 3/2)v)\frac{\dd R_{2S} }{\dd v} \nonumber \\
&- (2 A_S B_S + B_S/2 + A_S - 1/16 - F -H)  R_{2S} = 0 .
\end{align} 
This equation is of hypergeometric type (see Eq.\ (\ref{e: hypergeometric equation})) with parameters 
\begin{align}
 \alpha_S &= A_S + B_S + \tfrac{1}{4} + \tfrac{1}{2}\sqrt{\tfrac{k_S^2 + p}{p}}, \quad  \beta_S = A_S + B_S + \tfrac{1}{4} - \tfrac{1}{2}\sqrt{\tfrac{k_S^2 + p}{p}} , \nonumber \\ 
 \gamma_S &= 2 B_S + 1.
\end{align} 

In what follows we choose $A_S = 1/ 4$, $B_S = i \omega q / 2$, and we notice that the  parameters $\alpha_S, \beta_S, \gamma_S$ of the hypergeometric function fulfill $\gamma_S  -\beta_S -  \alpha_S = 0 $. From these results we obtain that the radial function $R_S$ is equal to \cite{Abramowitz-book}--\cite{NIST-book}
\begin{align} \label{e: solution scalar type}
 R_S=(1-v)^{1/4} v^{i \omega q/2} & \left(K_3 \,\,\,\, \hypergf (\alpha_S, \beta_S;\gamma_S;v) \right.  \\
 &+ \left. K_4 v^{1-\gamma_S} \hypergf (\alpha_S - \gamma_S +1,\beta_S - \gamma_S +1; 2 -\gamma_S; v ) \right) , \nonumber
\end{align}
where $K_3$ and $K_4$ are constants. Near the event horizon ($v=0$) of the topological black holes (\ref{e: 5D black holes})  this function behaves in a similar way to Eq.\ (\ref{e: near horizon vector type}). To fulfill the boundary condition i) of the QNM we must take $K_3 = 0$ and the function $R_S$ that satisfies this boundary condition is
\begin{align} \label{e: solution scalar type i)}
 R_S &= K_4 (1-v)^{1/4} v^{-i \omega q/2} \hypergf (\alpha_S - \gamma_S +1,\beta_S - \gamma_S +1; 2 -\gamma_S; v ) \nonumber \\
 &=  K_4 (1-v)^{1/4} v^{-i \omega q/2}  \hypergf (\hat{\alpha}_S, \hat{\beta}_S;\hat{\gamma}_S;v), 
\end{align} 
with
\begin{equation}
 \hat{\alpha}_S =  \alpha_S - \gamma_S +1, \qquad \hat{\beta}_S = \beta_S - \gamma_S +1, \qquad \hat{\gamma}_S = 2 -\gamma_S.
\end{equation} 

It is convenient to notice that the new parameters $\hat{\alpha}_S, \hat{\beta}_S, \hat{\gamma}_S$ also fulfill $\hat{\gamma}_S - \hat{\alpha}_S - \hat{\beta}_S = 0$. As for the vector type electromagnetic field, due to this fact we cannot use the Kummer property of the hypergeometric function \cite{Abramowitz-book}--\cite{NIST-book} to analyze the behavior of the scalar type electromagnetic field as $r \to + \infty$ ($v \to 1$). Taking into account the property (\ref{e: hypergeometric property integer}) of the hypergeometric function we find that the function $R_S$ satisfying the boundary condition near the horizon takes the form  
\begin{align}
  R_S &= K_4 v^{-i \omega q/2} (1-v)^{1/4} \frac{(-1)\Gamma(\hat{\gamma}_S) }{\Gamma(\hat{\alpha}_S)\Gamma(\hat{\beta}_S)} \sum_{s=0}^\infty \frac{(\hat{\alpha}_S)_s (\hat{\beta}_S)_s}{(s!)^2} (1-v)^s \nonumber \\
  & \times [\ln(1-v) -2\psi(s+1)  +\psi(\hat{\alpha}_S+s)+\psi(\hat{\beta}_S+s)]  
\end{align}
in the variable $1 -v$. As $r \to + \infty$ ($v \to 1$) this function behaves as 
\begin{align} \label{e: leading term scalar}
 R_S &\approx  (1-v)^{1/4} [\ln(1-v) -2\psi(1)  +\psi(\hat{\alpha}_S)+\psi(\hat{\beta}_S)]  \nonumber \\
 & \approx  \frac{L_1}{r^{1/2}} + \frac{L_2 \ln(r)}{r^{1/2}},
\end{align}
where $L_1$ and $L_2$ are constants. We notice that in the previous formula both terms go to zero as $r \to +\infty$. Thus the function $R_S$ of the expression (\ref{e: solution scalar type i)}) that fulfills the boundary condition near the horizon,  for all the frequencies, it also satisfies the boundary condition ii) of the QNM, thus, we shall obtain a continuous spectrum of QNF for the scalar type electromagnetic field (see for example Refs.\ \cite{Estrada-Jimenez:2013lra}, \cite{Stetsko:2016pau}). Nevertheless we expect to obtain a discrete set of quasinormal frequencies determined by the physical parameters of the black hole and the field \cite{Kokkotas:1999bd}, \cite{Berti:2009kk}. Therefore we must make a careful analysis of the behavior of the radial function $R_S$ as $r \to \infty$ before we impose the boundary condition ii) of the QNM. 

We find a similar example in the exact calculation of the QNF for the electromagnetic field propagating on an asymptotically Lifshitz black hole \cite{Lopez-Ortega:2014oha}. In the previous reference, to calculate the QNF of the electromagnetic field the boundary condition in the asymptotic region was modified to get a discrete spectrum of QNF. We see that in Ref.\ \cite{Lopez-Ortega:2014oha} it is proposed that for calculating the QNF of the electromagnetic field in the cases where the function $R_S$ is well behaved as $r \to + \infty$ we must impose as a boundary condition that the leading term of the asymptotic behavior must be canceled.

Following Ref.\ \cite{Lopez-Ortega:2014oha} to calculate the QNF of the scalar type electromagnetic field in the topological black holes (\ref{e: 5D black holes}) we propose that instead of the boundary condition ii) of the previous section, in the asymptotic region we must impose as boundary condition that the leading term in the asymptotic behavior (\ref{e: leading term scalar}) of the radial function $R_S$ must be canceled. We notice that for the vector type electromagnetic field when we calculate its QNF in the previous section we also cancel the leading term in the asymptotic behavior of the radial function $R_V$, but for the vector type electromagnetic field the leading term is divergent as $r \to +\infty$. Thus for the vector type electromagnetic field the new boundary condition as $r \to + \infty$ also can be used to compute its QNF and we obtain the same results of Sect.\ \ref{s: vector type}. We think that the new boundary condition that we impose at the asymptotic region is a natural generalization of the boundary condition ii).

In a similar way to Ref.\ \cite{Lopez-Ortega:2014oha} and motivated by the results of the previous section we assume that the parameters $\hat{\alpha}_S$ and $\hat{\beta}_S$ take the values
\begin{equation} \label{e: conditions QNF scalar type}
 \hat{\alpha}_S = -n, \qquad \qquad \qquad \hat{\beta}_S = -n,  \qquad \qquad \qquad n=0,1,2,\dots,
\end{equation} 
and with these values of the parameters we verify whether the radial function $R_S$ of the formula (\ref{e: solution scalar type i)}) satisfies the new boundary condition of the QNM as $r \to + \infty$. Thus taking $\hat{\alpha}_S = -n$ we get that this radial function becomes
\begin{equation} \label{e: radial scalar new boundary}
  R_S = K_4 v^{-i \omega q/2}  (1-v)^{1/4} \hypergf (-n,\hat{\beta}_S;\hat{\gamma}_S;v ).
\end{equation} 
Using the property (\ref{e: hypergeometric polynomial}) of the hypergeometric function we find that as $r \to + \infty$ the radial function (\ref{e: radial scalar new boundary}) behaves as
\begin{equation}
 R_S \approx \left( \frac{3 p}{\Lambda} \right)^{1/4} \frac{1}{r^{1/2}},
\end{equation} 
that is, we cancel the leading term of the asymptotic expansion (\ref{e: leading term scalar}) for $R_S$ and as $r \to + \infty$ the radial function (\ref{e: radial scalar new boundary}) fulfills  the new boundary condition of the QNM. Something similar happens when we take $\hat{\beta}_S = -n$. Therefore the QNF of the scalar type electromagnetic field are determined by the conditions (\ref{e: conditions QNF scalar type}) and they are equal to 
\begin{equation} \label{e: QNF scalar type}
 \omega_S = \pm \frac{1}{q} \sqrt{\frac{k_S^2-|p|}{|p|}} -i \frac{2}{q} \left( n + \frac{1}{2} \right) .
\end{equation} 

In contrast to the QNF of the vector type electromagnetic field, the previous QNF do not depend on the parameter $K$. Recalling that for $K=0, \pm 1$ the eigenvalues $k_S^2$ are non negative \cite{Kodama:2003kk}, for the three values of $K$ and for $k_S^2 > |p|$ we have complex QNF that are stable since $\im (\omega_S) < 0$. For   $k_S^2 < |p|$ the QNF (\ref{e: QNF scalar type}) transform into 
\begin{equation} \label{e: QNF scalar type eigenvalue small}
 \omega_S = \pm \frac{i}{q} \sqrt{\frac{|p| - k_S^2}{|p|}} -i \frac{2}{q} \left( n + \frac{1}{2} \right) ,
\end{equation} 
that are purely imaginary. Thus depending on the values of $|p|$ and $k_S^2$ we get complex or purely imaginary QNF for the scalar type electromagnetic field. For the minus sign of the square root in the formula (\ref{e: QNF scalar type eigenvalue small}) we have stable QNM, but for the plus sign of the square root we obtain QNF with $\im (\omega_S) > 0$ when we fulfill the following inequality 
\begin{equation}
 |p| - k_S^2 > 4 |p|\left( n + \frac{1}{2} \right)^2.
\end{equation} 
In a straightforward way we verify that this inequality cannot be satisfied for the allowed values of the physical parameters, that is, for $k_S^2 < |p|$ also we obtain stable QNM. Thus for the scalar type electromagnetic field we find only stable QNM in the five dimensional topological black holes (\ref{e: 5D black holes}).

\section{Discussion}
\label{s: Discussion}

In the previous sections we calculate exactly the QNF of the vector type and the scalar type electromagnetic fields propagating on the topological black holes (\ref{e: 5D black holes}). It is convenient to notice that for the three values of the parameter $K$ we state the radial problem in a common form and we solve simultaneously the differential equations. It is convenient to notice that the method previously used to calculate the QNF (\ref{e: QNF vector type}) and (\ref{e: QNF scalar type}) is slightly different from the used in other references, since the special values of the parameters for the hypergeometric functions that we obtain in the topological black hole (\ref{e: 5D black holes}), force us to impose the conditions (\ref{e: conditions vector type}) and (\ref{e: conditions QNF scalar type}) and then verify that for these values of the parameters the radial functions satisfy the boundary condition at infinity. Usually it possible to employ the Kummer property of the hypergeometric function and to choose the appropriate behavior at the boundaries by imposing the analogue of the conditions (\ref{e: conditions vector type}) and (\ref{e: conditions QNF scalar type}) \cite{Cardoso:2001hn}--\cite{Becar:2012bj}. 

Depending on the physical parameters for the scalar type and vector type electromagnetic fields we obtain complex QNF or purely imaginary QNF. We find that the QNM of the electromagnetic field are stable, except for the topological black holes with $K=-1$ of the plus branch for which we find that for small values of the parameter $|p|$ and of the eigenvalues $k_V^2$ for the vector harmonics, the QNM of the vector type electromagnetic field would be unstable if the three-dimensional base manifold has eigenvalues $k_V^2$ satisfying $0 < k_V^2 < 2$. Therefore, as noted previously for other backgrounds in Refs.\ \cite{Kodama:2003kk}, \cite{Gibbons:2002pq}, the eigenvalues of the scalar and vector harmonics of the base manifold play a relevant role in the analysis of the classical stability of the black holes under perturbations. 

Comparing the expressions (\ref{e: QNF vector type}) and (\ref{e: QNF scalar type}) for the QNF of the vector type and scalar type electromagnetic fields, we see that they are not isospectral since the terms in square roots show a different dependence on the parameter $p$. Also, notice that the QNF of the vector type electromagnetic field depends on the scalar curvature $K$, whereas the QNF of the scalar type electromagnetic field are independent of $K$. Furthermore the QNF of the vector type field depend on the overtone number $n$ in the form $(n+1)$, whereas the QNF of the scalar type field depend on $n$ in the form $(n+1/2)$. 

Finally considering that the Hawking temperature of the topological black holes (\ref{e: 5D black holes}) is \cite{Cai:2009ar}
\begin{equation}
 T_H = \frac{1}{2 \pi q} ,
\end{equation} 
for the vector type electromagnetic field we can write its QNF (\ref{e: QNF vector type}) as
\begin{equation}
 \omega_V = \pm 2 \pi T_H \sqrt{\frac{k_V^2 + 2  K}{|p|}} - i 4 \pi T_H (n + 1) ,
\end{equation} 
and for the scalar type electromagnetic field its QNF (\ref{e: QNF scalar type}) as 
\begin{equation}
 \omega_S =  \pm 2 \pi T_H \sqrt{ \frac{k_S^2 - |p|}{|p|} }  - i 4 \pi T_H  \left( n + \frac{1}{2} \right)  .
\end{equation} 
Thus for the five-dimensional topological black holes (\ref{e: 5D black holes}) the QNF are proportional to its Hawking temperature.

\section{Acknowledgments}

This work was supported by CONACYT M\'exico, SNI M\'exico, EDI-IPN, COFAA-IPN, and Research Project IPN SIP-20171817.


\begin{thebibliography}{}

   
\bibitem{Kokkotas:1999bd}
  K.~D.~Kokkotas and B.~G.~Schmidt,
  Living Rev.\ Rel.\  {\bf 2}, 2 (1999)
  [arXiv:gr-qc/9909058].

\bibitem{Berti:2009kk}
  E.~Berti, V.~Cardoso and A.~O.~Starinets,
  Class.\ Quant.\ Grav.\  {\bf 26}, 163001 (2009)
  [arXiv:0905.2975 [gr-qc]].
  
\bibitem{Hod:1998vk}
  S.~Hod,
  Phys.\ Rev.\ Lett.\  {\bf 81}, 4293 (1998)
  [arXiv:gr-qc/9812002].

\bibitem{Maggiore:2007nq}
  M.~Maggiore,
  Phys.\ Rev.\ Lett.\  {\bf 100}, 141301 (2008)
  [arXiv:0711.3145 [gr-qc]].
  
\bibitem{Aharony:1999ti} 
  O.~Aharony, S.~S.~Gubser, J.~M.~Maldacena, H.~Ooguri and Y.~Oz,
  Phys.\ Rept.\  {\bf 323}, 183 (2000)
  [hep-th/9905111].

\bibitem{Chan:1996yk} 
  J.~S.~F.~Chan and R.~B.~Mann,
  Phys.\ Rev.\ D {\bf 55}, 7546 (1997)
  [gr-qc/9612026].

\bibitem{Horowitz:1999jd} 
  G.~T.~Horowitz and V.~E.~Hubeny,
  Phys.\ Rev.\ D {\bf 62}, 024027 (2000)
  [hep-th/9909056].
 
\bibitem{Cardoso:2001bb} 
  V.~Cardoso and J.~P.~S.~Lemos,
  Phys.\ Rev.\ D {\bf 64}, 084017 (2001)
  [gr-qc/0105103].
 
\bibitem{Cardoso:2001vs} 
  V.~Cardoso and J.~P.~S.~Lemos,
  Class.\ Quant.\ Grav.\  {\bf 18}, 5257 (2001)
  [gr-qc/0107098].
  
\bibitem{Cardoso:2003cj} 
  V.~Cardoso, R.~Konoplya and J.~P.~S.~Lemos,
  Phys.\ Rev.\ D {\bf 68}, 044024 (2003)
  [gr-qc/0305037].
  
\bibitem{Berti:2003ud} 
  E.~Berti and K.~D.~Kokkotas,
  Phys.\ Rev.\ D {\bf 67}, 064020 (2003)
  [gr-qc/0301052].

\bibitem{LopezOrtega:2006vn} 
  A.~Lopez-Ortega,
  Gen.\ Rel.\ Grav.\  {\bf 38}, 1747 (2006)
  [gr-qc/0605034].
  
\bibitem{Starinets:2002br} 
  A.~O.~Starinets,
  Phys.\ Rev.\ D {\bf 66}, 124013 (2002)
  [hep-th/0207133].
  
\bibitem{Nunez:2003eq} 
  A.~Nunez and A.~O.~Starinets,
  Phys.\ Rev.\ D {\bf 67}, 124013 (2003)
  [hep-th/0302026].
 
\bibitem{Kovtun:2005ev} 
  P.~K.~Kovtun and A.~O.~Starinets,
  Phys.\ Rev.\ D {\bf 72}, 086009 (2005)
  [hep-th/0506184].
 
\bibitem{Musiri:2005ev} 
  S.~Musiri, S.~Ness and G.~Siopsis,
  Phys.\ Rev.\ D {\bf 73}, 064001 (2006)
  [hep-th/0511113].

\bibitem{Miranda:2008vb} 
  A.~S.~Miranda, J.~Morgan and V.~T.~Zanchin,
  JHEP {\bf 0811}, 030 (2008)
  [arXiv:0809.0297 [hep-th]].
  
\bibitem{Cardoso:2001hn}
  V.~Cardoso and J.~P.~S.~Lemos,
  Phys.\ Rev.\  D {\bf 63}, 124015 (2001)
  [arXiv:gr-qc/0101052].

\bibitem{Cordero:2012je} 
  R.~Cordero, A.~Lopez-Ortega and I.~Vega-Acevedo,
  Gen.\ Rel.\ Grav.\  {\bf 44}, 917 (2012)
  [arXiv:1201.3605 [gr-qc]].
  
\bibitem{Birmingham:2001pj}
  D.~Birmingham, I.~Sachs and S.~N.~Solodukhin,
  Phys.\ Rev.\ Lett.\  {\bf 88}, 151301 (2002)
  [arXiv:hep-th/0112055].
  
\bibitem{Aros:2002te}
  R.~Aros, C.~Martinez, R.~Troncoso and J.~Zanelli,
  Phys.\ Rev.\  D {\bf 67}, 044014 (2003)
  [arXiv:hep-th/0211024].

\bibitem{Birmingham:2006zx}
  D.~Birmingham and S.~Mokhtari,
  Phys.\ Rev.\  D {\bf 74}, 084026 (2006)
  [arXiv:hep-th/0609028].

\bibitem{LopezOrtega:2007vu}
  A.~Lopez-Ortega,
  Gen.\ Rel.\ Grav.\  {\bf 40}, 1379 (2008)
  [arXiv:0706.2933 [gr-qc]].
  
\bibitem{LopezOrtega:2010uu} 
  A.~Lopez-Ortega,
  Rev.\ Mex.\ Fis.\  {\bf 56}, 44 (2010)
  [arXiv:1006.4906 [gr-qc]].
  
\bibitem{Becar:2012bj} 
  R.~Becar, P.~A.~Gonzalez and Y.~Vasquez,
  Int.\ J.\ Mod.\ Phys.\ D {\bf 22}, 1350007 (2013)
  [arXiv:1210.7561 [gr-qc]].

\bibitem{Cai:1998vy} 
  R.~G.~Cai and K.~S.~Soh,
  Phys.\ Rev.\ D {\bf 59}, 044013 (1999)
  [gr-qc/9808067].
  
\bibitem{Cai:2001dz} 
  R.~G.~Cai,
  Phys.\ Rev.\ D {\bf 65}, 084014 (2002)
  [hep-th/0109133].

\bibitem{Cai:2009ar} 
  R.~G.~Cai, Y.~Liu and Y.~W.~Sun,
  JHEP {\bf 0906}, 010 (2009)
  [arXiv:0904.4104 [hep-th]].
  
\bibitem{Gonzalez:2010vv} 
  P.~Gonzalez, E.~Papantonopoulos and J.~Saavedra,
  JHEP {\bf 1008}, 050 (2010)
  [arXiv:1003.1381 [hep-th]].

\bibitem{Becar:2013qba} 
  R.~Becar, P.~A.~Gonzalez and Y.~Vasquez,
  Phys.\ Rev.\ D {\bf 89}, 023001 (2014)
  [arXiv:1306.5974 [gr-qc]].
  
\bibitem{Oliva:2010xn} 
  J.~Oliva and R.~Troncoso,
  Phys.\ Rev.\ D {\bf 82}, 027502 (2010)
  [arXiv:1003.2256 [hep-th]].
  
\bibitem{Kodama:2003kk} 
  H.~Kodama and A.~Ishibashi,
  Prog.\ Theor.\ Phys.\  {\bf 111}, 29 (2004)
  [hep-th/0308128].

\bibitem{Lopez-Ortega:2014oha} 
  A.~Lopez-Ortega,
  Gen.\ Rel.\ Grav.\  {\bf 46}, 1756 (2014)
  [arXiv:1406.0126 [gr-qc]].
 
\bibitem{Abramowitz-book}  M.~Abramowitz and I.~A.~Stegun, {\it Handbook of Mathematical Functions, Graphs, and Mathematical Table}, (Dover Publications, New York, 1965).

\bibitem{Guo-book} Z.~X.~Wang and  D.~R.~Guo, {\it Special Functions}, (World Scientific Publishing, Singapore, 1989).

\bibitem{NIST-book} F.~W.~J.~Olver, D.~W.~Lozier, R.~F.~Boisvert, and C.~W.~Clark, {\it NIST Handbook of Mathematical Functions}, (Cambridge University Press, New York, 2010).

\bibitem{Estrada-Jimenez:2013lra} 
  S.~Estrada-Jimenez, J.~R.~Gomez-Diaz and A.~Lopez-Ortega,
  Gen.\ Rel.\ Grav.\  {\bf 45}, 2239 (2013)
  [arXiv:1308.5943 [gr-qc]].

\bibitem{Stetsko:2016pau} 
  M.~M.~Stetsko,
  arXiv:1612.09172 [hep-th].

\bibitem{Gibbons:2002pq} 
  G.~Gibbons and S.~A.~Hartnoll,
  Phys.\ Rev.\ D {\bf 66}, 064024 (2002)
  [hep-th/0206202].
  


\end{thebibliography}
\end{document}